 \documentstyle[aps,prd,epsfig,preprint,floats]{revtex} 

\tightenlines	
\input epsf.tex
\def\DESepsf(#1 width #2){\epsfxsize=#2 \epsfbox{#1}}
\newcommand{\ifm}[1]{\relax\ifmmode #1\else $#1$\fi}
 
\newcommand{\beq   }{\begin{equation}}
\newcommand{\eeq   }{\end{equation}}
\newcommand{\beqn  }{\begin{eqnarray}}
\newcommand{\eeqn  }{\end{eqnarray}}
\newcommand{\bi    }{\begin{itemize}}
\newcommand{\ei    }{\end{itemize}}
\newcommand{\bc    }{\begin{center}}
\newcommand{\ec    }{\end{center}}
\newcommand{\bd    }{\begin{description}}
\newcommand{\ed    }{\end{description}}
\newcommand{\bHuge }{\begin{Huge}}
\newcommand{\bhuge }{\begin{huge}}
\newcommand{\bLARGE}{\begin{LARGE}}
\newcommand{\bLarge}{\begin{Large}}
\newcommand{\blarge}{\begin{large}}
\newcommand{\eHuge }{\end{Huge}}
\newcommand{\ehuge }{\end{huge}}
\newcommand{\eLARGE}{\end{LARGE}}
\newcommand{\eLarge}{\end{Large}}
\newcommand{\elarge}{\end{large}}

\def \gtsim    {\relax\ifmmode{\mathrel{\mathpalette\oversim >}}
                  \else{$\mathrel{\mathpalette\oversim >}$}\fi}
\def \ltsim    {\relax\ifmmode{\mathrel{\mathpalette\oversim <}}
                  \else{$\mathrel{\mathpalette\oversim <}$}\fi}
\def\oversim#1#2{\lower4pt\vbox{\baselineskip0pt \lineskip1.5pt
            \ialign{$\mathsurround=0pt#1\hfil##\hfil$\crcr#2\crcr\sim\crcr}}}

\newcommand{\gev}  {\mbox{${\rm GeV}$}}

\newcommand{\gevcc}{\mbox{${\rm GeV}/c^2$}}

\newcommand{\tev}  {\mbox{${\rm TeV}$}}

\newcommand{\invpb}{\mbox{${\rm pb}^{-1}$}}

\newcommand{\invfb}{\mbox{${\rm fb}^{-1}$}}

\newcommand{\bbar} {\mbox{$\overline{b}$}}

\newcommand{\mpmm} {\mbox{$\mu^+\mu^-$}}
\newcommand{\azero}{\ifm{A_0}}
\newcommand{\tanb}{\ifm{\tan\beta}}
\newcommand{\mzero}{\ifm{m_0}}
\newcommand{\mhalf}{\ifm{m_{1/2}}}
\newcommand{\rpv} {\mbox{${R\!\!\!\!\!\:/_p}$}}

\newcommand{ \stau}     {\mbox{$\tilde{\tau}$}}

\newcommand{ \sstop}    {\mbox{$\tilde{t}$}}

\newcommand{ \schi }    {\mbox{$\tilde{\chi}$}}
\newcommand{ \lsp}      {\mbox{$\tilde{\chi}_{1}^{0}$}}

\newcommand{ \bsmumu }{\mbox{$B_{s} \to \mu^+ \mu^-$}}
\newcommand{ \bsgam }{\mbox{$b \to s \gamma$}}

\def \etal     {\relax\ifmmode{et \; al.}\else{$et \; al.$}\fi}
\newcommand{\ie}{$i.e.$}
\newcommand{\eg}{$e.g.$}
\newcommand{\eps}{\mbox{$\epsilon$}}
\def \calR     {\relax\ifmmode{{\cal R}}\else{${\cal R}$}\fi}

\begin{document}
\draft
\preprint{\hbox{CTP-TAMU-02-02}}
\title{
\boldmath  Detection of  \bsmumu\
 at the Tevatron Run II and Constraints on the SUSY Parameter Space}

\author{ R. Arnowitt$^\ast$, B. Dutta$^\ast$, T. Kamon$^\ast$,\\
M. Tanaka$^\dag$\\[.2in]
$^\ast$Department of Physics, Texas A\&M University,  College Station TX
77843-4242,\\ 
$^\dag$Argonne National Laboratory, Argonne, IL 60439.}
\date{\today}
\maketitle


\begin{abstract} A measurement of the branching ratio for the rare decay mode
\bsmumu\ at  the Tevatron is an opportunity to test various supersymmetric
scenarios. We  investigate the prospects for studying this mode in Run II and
estimate  that CDF would be sensitive to this decay for a branching ratio 
$> 1.2  \times 10^{-8}$ with 15 \invfb (or, if a similar analysis holds for D0,
$>6.5\times 10^{-9}$ for the combined data).  We calculate the branching ratio in
minimal supergravity (mSUGRA) parameter space, and find that $\tanb > 30$ can be probed. (This mSUGRA 
parameter space cannot be probed by direct production of SUSY particles at  Run
II.) Including other experimental constraints on the mSUGRA parameter  space,
one finds that CDF \bsmumu\ measurements would be able to cover  the full mSUGRA
parameter space for $\tanb = 50$ if the muon $g_{\mu} - 2$ anomaly  exceeds $\sim 11
\times 10^{-10}$,  and about half the allowed parameter space for 
\tanb\ = 40. A large branching ratio $> 7(14) \times 10^{-8}$  (feasible with only 
2 \invfb) would be sufficient to exclude the mSUGRA model for
$\tan\beta\leq50(55)$. Dark matter 
neutralino-proton detection cross sections are examined in the allowed  region,
and should be large enough to be accessible to future planned  experiments.
Combined measurements of $B_s\rightarrow\mu^+\mu^-$, the Higgs mass $m_h$ and
the muon $g_{\mu}-2$ anomaly would be sufficient to determine the $\mu>0$ mSUGRA
parameters (or show the model is inconsistent with the data). We
also briefly discuss the \bsmumu\  decay in R parity  violating models. There,
for some models, the branching ratio can be large  enough to be detected even
for small $\tan\beta$ and large \mhalf.\\
\\ 

\end{abstract}
 \pacs{ }   



\noindent {\bf 1. INTRODUCTION}.  Supersymmetric (SUSY) extensions of the Standard
Model (SM) represent a  natural candidate for the new physics expected to occur
in the TeV energy  domain. One of the difficulties in determining predictions of
such models  lies in the large number of new parameters the theory implies. Thus
the  most general low energy model,  the Minimal Supersymmetric Standard Model
(MSSM),
 has over 100 free parameters, and  even the minimal model based on supergravity
grand unification at the scale 
$M_G = 2 \times 10^{16}\ \gev$ 
 (mSUGRA) possesses four new parameters and one sign  in addition to the SM
parameters.  Fortunately SUSY models apply to a large  number of different
accelerator and cosmological phenomena, and a great  deal of effort has been
involved in recent years to to use this data to  limit the parameter space. Part
of the difficulty in doing this resides in  the success of the model in not
disturbing the excellent agreement of the  precision tests of the SM 
\cite{Langacker:2001ij} due to
the SUSY decoupling theorems which  suppress SUSY contributions at low energies.
Historically, the absence of  flavor changing neutral currents at the tree level
played an important role  in the construction of the SM. They represent
therefore an important class  of phenomena that might show the presence of new
physics, since the SM and  the SUSY contributions contribute first at the loop
level with comparable  size.  Thus the decay \bsgam\ has been a powerful tool in
limiting  the SUSY parameter space. In this paper, we consider the decay \bsmumu\
within the framework of mSUGRA models and R parity violating models. This process is particularly  interesting
for several reasons: The SM branching ratio is quite small,
\ie,   $Br[\bsmumu]_{\rm SM} = 3.5 \times 10^{-9}$ \cite{Anikeev:2001rk}.  The SUSY contribution
\cite{Babu:1999hn,Chankowski:uz,Bobeth:2001sq,Dedes:2001fv,Isidori:2001fv,Huang:2002dj} has terms that grow as $\tan^{6}\beta$
 (where $\tan\beta = <H_2> / <H_1>$  and $H_{1,2}$ are the two SUSY Higgs
bosons) and thus can become quite large  for large \tanb. Finally, as we shall
show below, the two collider  detectors at the Tevatron will  be sensitive to this
decay for
$\tan\beta\gtsim\ 30$ in Run II.

\noindent {\bf 2. MSUGRA MODELS}. We consider first mSUGRA models with R parity invariance, and
combine the  existing accelerator constraints on the parameter space (\eg, the
light  Higgs mass $m_h$ bound, \bsgam\, $etc.$),  the cosmological dark matter 
constraints, the possible muon magnetic moment anomaly, with what might  be
expected from the Tevatron measurement of  the \bsmumu\ decay. The  combined
constraints can significantly limit the SUSY parameter space, and  thus allow
better predictions as to what the models predict at the LHC.

The mSUGRA model \cite{Chamseddine:jx,sugra2} depends on four parameters which we take to be  the
following:
\mzero\ (the universal scalar mass at $M_G$), 
\mhalf\ (the universal gaugino mass at $M_G$), 
\azero\  (the universal cubic soft breaking mass at $M_G$),  and $\tanb$ (at the
electroweak scale).  In addition, the sign of $\mu$ (the  Higgs mixing parameter
appearing in the superpotential as $\mu H_1 H_2$) is  arbitrary.  We consider
the parameter range of 
$\mzero, \mhalf < 1\ \tev$, 
$3 < \tanb < 55$, and $| \azero | < 4 \mhalf$.  We take a $2\sigma$ bound on the
\bsgam\ decay \cite{Chen:2001fj} of 
$1.8 \times  10^{-4} < Br[\bsgam] < 4.5 \times 10^{-4}$,  and the  LEP bound on
the light Higgs of $m_h > 113.5\ \gev$\cite{higgs}.  Since there is still  a (2-3) GeV
uncertainty in the theoretical calculation of $m_h$, we will  (conservatively)
interpret this to mean $m_h {\rm (theory)} > 111\ \gev$.  For 
$\tanb\ \gtsim\ 45$, results are sensitive to the precise values of
$m_b$ and $m_t$.  We assume here that $m_b(m_b)$ = 4.25 GeV and $m_t(pole)$ =
175 GeV, and  include the large $\tan\beta$ corrections to the $b$-quark Yukawa
coupling  constant \cite{Rattazzi:1995gk}. We assume that any  muon $g - 2$ deviation from the SM
is due  to SUSY \cite{Yuan:ww,Kosower:1983yw}.  The experimental deviation has now been reduced to a
$1.6\, \sigma$  effect \cite{Knecht:2001qg,Hayakawa:2001bb}, and we take here this deviation to be greater
than $1 \sigma$  below the current central value 
\ie,  $a_{\mu}^{\rm SUGRA} > 11 \times 10^{-10}$.  We will  show however, what
would happen if this deviation were  to change with the  new BNL E821 data
currently being analyzed. For models with R parity  invariance, the lightest
neutralino, \lsp\ is the dark matter particle,  and we require that $0.07 <
\Omega_{\lsp} h^2 < 0.21$,  in accord with current CMB  and other astronomical
data \cite{Turner:2002fi}. All stau-neutralino co-annihilation  channels
\cite{Arnowitt:2001yh,ellis,gomez} are included in
the relic density calculations.

\noindent {\bf 3. ${\boldmath \bsmumu}$ DECAY}.  The branching ratio for $\bsmumu$ is given in
\cite{Bobeth:2001sq} which we write in the form
\beqn  Br[\bsmumu] & = &  {{2\tau_B
M_B^5}\over{64\pi}}f^2_{B_s}\sqrt{1-{{4m_l^2}\over{M_B^2}}} 
	\nonumber \\
	& &
 \left[ \left( 1-{{4m_l^2}\over{M_B^2}} \right) 
	\left| {{(C_S-C_S')}\over{(m_b+m_s)}} \right|^2+
 \left| {{(C_P-C_P')}\over{(m_b+m_s)}}+
	2{m_{\mu}\over M_{B_s}^2}(C_A-C_A') \right|^2
	\right]~~
\label{eq:bsmm_br_msugra}
\eeqn
\par\noindent where $f_{B_s}$  is the $B_s$ decay constant,  
$M_{B}$ is
the $B$ meson mass, $\tau_B$ is the mean life and $m_l$ is the mass of lepton.
$C_S$, $C^{\prime}_S$, $C_P$, 
$C^{\prime}_P$ include the SUSY loop contributions  due to diagrams involving
the  particles such as stop, chargino, sneutrino, Higgs etc..  For large $\tan\beta$,
the  amplitude has terms that grow like $\tan^{3}\beta$  as can be seen in the
example of Figure \ref{fig1}.  Thus at large $\tan\beta$, the dominant contribution to
$C_S$ is given approximately by
\beqn
 C_S &\simeq & 
 {{G_F\alpha}\over {\sqrt 2\pi}}V_{tb}V_{ts}^* 
 \left( {\tan^3\beta\over{4\sin^2\theta_W}} \right) 
 \left({{m_bm_{\mu} m_t \mu}\over{M_W^2 M_A^2}} \right) 
 {\sin2\theta_{\tilde t}\over 2} \left( \frac{ m_{\tilde t_1}^2
\log\left[\frac{m_{\tilde t_1}^2}{\mu^2}\right] }
	{ \mu^2-m_{\tilde t_1}^2 } -
  \frac{ m_{\tilde t_2}^2 \log\left[\frac{m_{\tilde t_2}^2}{\mu^2}\right] }
	{ \mu^2-m_{\tilde t_2}^2 } \right)
\eeqn 
\par\noindent where $m_{\sstop_{1,2}}$ are the two stop masses,  and
$\theta_{\sstop}$  is the rotation angle  to diagonalize the stop mass matrix.
We need to muliply  the above expression by
$1/(1+\epsilon_b)^2$ to include the SUSY QCD corrections. $\epsilon_b$ is
proportional to  $\mu\tan\beta$ \cite{Carena:1999py}.  We have $C_P = -C_S,  C^{\prime}_S = 
(m_s/m_b)C_s$  and $C^{\prime}_P = - (m_s/m_b)C_P$. The operators are given in
ref.\cite{Bobeth:2001sq}. In our numerical calculation, we use all complete one loop
contributions to the branching ratio.

\noindent {\bf 4. DETECTION OF ${\boldmath \bsmumu}$ at TEVATRON}. We consider now the
possibility of detecting the decay \bsmumu\ by  the CDF and D0 detectors at the
Tevatron in Run II. Both detectors have been upgraded with excellent tracking and
muon detector systems \cite{Anikeev:2001rk}. The dimuon trigger is the key to collect the 
$B_{s} \to \mu^+\mu^-$ decays.

In order to estimate the limits on $Br[B_s \to \mpmm]$ detection,  we use the 95$\%$
C.L. limit on $Br[B_s \to \mpmm]$ published by CDF\cite{Abe:1998ah}. Thus our discussion is based
on the CDF detector, although both CDF and D0 detectors should have a similar
perfromance.

In the Run I analysis, CDF observed one candidate that was
consistent with $B_s \to \mpmm$ with an estimate of 0.9 backound (BG) events in 98 \invpb\
\cite{Abe:1998ah}. The primary Run-I selection variables and cut values were
that at
least one muon track with 
$c\tau \equiv L_{xy} M_{B} / P_{T}^{\mu\mu} >100\ \mu$m,  $I\equiv
P^{\mu\mu}_T/[P^{\mu\mu}_T+\Sigma P_T]>0.75$ for the muon pair, 
and $\Delta\Phi<0.1$ rad.  Here, $L_{xy}$ is the transverse decay
length;
$p_{T}^{\mu\mu}$ is the transverse momentum of the dimuon system. 
$\Sigma P_T$ is the scalar sum of the transverse momenta of
all tracks, excluding the muon candidates, within a cone of $\Delta
R\equiv\sqrt{(\Delta\eta)^2+(\Delta\phi)^2}=1$ around the monetum vector of the muon pair. 
The $z$ coordinate
of each track  along the beam line \cite{def} must be within 5 cm of the primary vertex. $\Delta\Phi$ is an
opening azimuthal angle between $P_T^{\mu\mu}$ and the vector pointing from the
primary vertex to the secondary vertex (the reconstructed $B$-meson decay
position). As a conservative estimate, CDF took the one event as signal
to calculate 95\% C.L. limit of signal events ($N_1^{95\%}\equiv$5.06
events\cite{Abe:1998ah}) and had set a limit of $Br <
2.6 \times 10^{-6}$. In the analysis,  the selection efficiency ($\eps$) for
signal events and the rejection power (${\cal R}$) for background events (pass a baseline
selection \cite{Abe:1998ah}) are
estimated to be $\eps_{1} = 0.45$ and $\calR_{1} = 440$ by using a sample of
like-sign dimuon events ($5 < M_{\mu\mu} < 6\ \gevcc$).


A dimuon trigger in Ref.\cite{Anikeev:2001rk} will improve the acceptance for
signal events by a factor of 2.8. The trigger will soon be tested using the Run
IIa (2 \invfb) data. This will allow us to modify the trigger design for
the higher luminosity expected in Run IIb (15 \invfb). 
In our analysis, we assume that the
dimuon trigger can be designed by maintaing the acceptance for signal events. 
We expect to improve the acceptance for signal events by a factor of
2.8 \cite{Anikeev:2001rk}. If we assume the factor 2.8 to be the same for BG
events, then we would observe 51 (386) events in  2 (15) \invfb\  with the same
cuts as in Run I. Therefore, CDF has to require a set of tighter cuts to obtain
the best possible upper limit.

We first need to understand the  background contents and expected improvement by
the  new Run II detector.  Two types of backgrounds must be taken into account:
(i) non-$b$ backgrounds comming from the primary vertex; (ii) $b$ background
events, such as the gluon-spliting
$b\bbar$ events.

The most important feature for $B$ decays is the displaced vertex. One way to
reduce prompt background is to require  a minimum decay length $L_{xy}$.
However, two tracks can appear to form a secondary vertex  if one of two tracks
originates from the primary vertex and the other has an impact parameter ($d$).
Therefore, the requirement of a minimal impact parameter of individual tracks
can further clean up the sample. It has been shown for example, in the Run-I analysis for
$B^0 \to K^{0*} \mu^+ \mu^-$ events \cite{Affolder:1999eb}, that a tight impact
parameter  cut
on significance for individual track ($d/\sigma_{d}>2$) significantly improve the
background rejection even with $L_{xy} > 100~\mu$m. One has then  $\eps \sim
\eps_{1} \times 0.43$ and 
$\calR \sim \calR_{1} \times 190$.

We indeed observed the similar numbers of OS and LS events in the Run-I final
selection for $B_{s} \to \mpmm$ events, while we should expect more OS events
than LS events  if the cuts were tight enough for non-$b$ background events.
This indicates the final dimuon event candidates in Run I analysis are actually dominated by the
non-$b$ background sources. Thus a higher 
track impact parameter is neccessary to reduce the non-$b$ backgrounds.
We would expect larger reduction with good efficiency even after the
$L_{xy}$ cut. In  other words, we would have been able to set  better limits  if
we had included the impact parameter for the cut optimization. The silicon
vertex detector (SVX-II) will
provide us much better reduction  for the non-$b$ background than Run I. The
non-$b$ background will not be a problem in Run II.

In Run II, the most severe background will be the two muons from gluon-spliting
$b\bbar$ events. Since both particles are $b$ quarks, the impact parameter does
not help. Both $b$ and $\bbar$ also  go in  the same direction,  so that cut on
$L_{xy}$ does not help either. However, $\Delta\Phi$ is still usefull to remove
the background events. Furthermore, in Run II, we can use $\Delta\Theta$ in
$r$-$z$ view since we have $z$-strips in SVX-II. 

There is some room to improve the isolation cut.  We can form a new isolation
parameter by only using  the tracks with large impact parameter. This new 
isolation cut will work to reject the $b\bbar$ rather than non-$b$ background.
Furthermore, we can search for tracks with large impact parameter on the
opposite side of the dimuon candidates to make sure that the $b$ and $\bbar$ go
to the opposite side.

Therefore, CDF could improve the BG rejection by a factor of 200-400 with
further reduction of the signal efficiency by a factor of 2-3.  Based on these
facts, we now consider two cases to evaluate  Run II limits as a function of
luminosity.

In the first case (Case A), we naively assume new tighter cuts in Run II,
described above, will gain additional BG rejection power of 450 for additonal
efficiency of 0.45, or
  $ {\cal R}_2  =  450^{0.45/\eps_2}$. This gives us
\beqn
  \frac{\eps_2}{\eps_1} & = & 
	\frac{1}{1 + \log({\cal R}_2/{\cal R}_1)/\log(450)}
\label{eq:eff_vs_BGrejection}
\eeqn If we could optimize the BG rejection in Run IIa (2 \invfb) to be
$\calR_2 \approx 51 \calR_1$ with $\eps_2 \approx 0.61 \eps_1$ (from Eq.
\ref{eq:eff_vs_BGrejection}), then we would expect  one BG event in 2 \invfb.
Thus, with an assumption of the same size of the  total systematic uncertainty in
Run II as in Run I, we can extrapolate the 95\%
C.L. limit to be 
$ Br < 7.7 \times 10^{-8}$ for 2 fb$^{-1}$ using $N_1^{95\%}$.

In the second case (Case B), we simply assume the Run-II background rejection
could be improved
	(without loosing the signal efficiency) 
	to keep the expected BG events in 2 \invfb\
	 at the level of Run I (\ie, 0.9 events).
	If we would observe one event in 2 \invfb,
	then
	we could set the limits by scaling the Run-I $Br$ limit down  by the
luminosity (2000 \invpb/98 \invpb) and the acceptance by (2.8/1.0). Thus we
obtain
$Br < 4.6 \times 10^{-8}$.
	This would certainly be the optimistic scenario, but
	it would be a goal of this analysis in Run IIa.
Here, the systematic uncertainty in Run II is assummed to be the same as in Run
I. 

We repeat the same arguement for different luminosities.
Figure \ref{fig:bsmumu_br_limits} shows 95\% C.L. limits on $Br[B_s \to \mpmm]$
at CDF in Run II as a function of integrated luminosity for Cases A and B. For
15 $\rm fb^{-1}$ in case A, CDF would be sensitive to $Br>1.2\times 10^{-8}$ and the combined
CDF and D0 data (30 $\rm fb^{-1}$) would be sensitive to $Br>6.5\times 10^{-9}$.

\noindent {\bf 5. RESULTS FOR MSUGRA}. We examine first the parameter region for the
mSUGRA model that would be  accessible to CDF at Run II with 15 \invfb\ of data. 
Figure \ref{fig:sugra1} shows the  
$Br[\bsmumu]$ as a function of \mhalf\ for \azero\ = 0, 
\mzero \,= 300 GeV. One sees  that with a sensitivity of $Br[\bsmumu] > 1.2 \times
10^{-8}$
 for 15 \invfb, the Tevatron Run II can probe the \bsmumu\  decay  for $\tanb\ >
30$.  Further, a search for this decay would sample much higher regions of 
\mhalf\  than a direct search at Run II for SUSY particles which is restricted
to 
$\mhalf < 250\ \gev$  \cite{Barger:1998hp}.  As \mzero\ increases, the branching ratio goes
down.  However, this dependence becomes less significant for large 
\mhalf, where 
\mzero\  as large as 800 GeV can be sampled for large \mhalf.

In Figure \ref{fig:sugra3}
 the contours of $Br[\bsmumu]$ are plotted in the 
\mzero-\mhalf\  plane for \tanb\ = 50, \azero\ = 0.  We combine now this result
with the other  experimental constraints. Thus the shaded region to the left is
ruled out  by the \bsgam\     constraint, and the shaded region on the right
hand side  is disallowed if $a_\mu^{\rm SUGRA} > 11 \times 10^{-10}$.  The
narrow shaded band in the  middle is allowed by the dark matter constraint. We
note that independent  of whether the astronomically observed dark matter is
SUSY in origin, the  dark matter allowed region for mSUGRA cannot significantly
deviate from  this shaded region, for below the narrow shaded band, the stau
would be  lighter than the neutralino (leading to charged dark matter), while
above  the band, mSUGRA would predict more neutralino dark matter than is
observed.

Using our estimate that $Br > 1.2 \times 10^{-8}$  can be observed with 15
\invfb,  we see that almost the entire parameter space allowed by the $a_\mu$ 
constraint can be probed in Run II for \tanb\ = 50. Note that an observed 
$Br[\bsmumu] >  7 \times 10^{-8}$, possible with only 2 \invfb (see Figure
\ref{fig:bsmumu_br_limits}),  would be sufficient to rule out the mSUGRA model
for $\tan\beta\leq 50$. 
In Figure \ref{fig:sugra3} we also show  the expected dark matter detector cross
section for Milky Way dark matter  (the short solid lines). They are of a size
that can be observed by future  planned dark matter detectors 
such as GENIUS, Cryoarray, ZEPLIN IV and CUORE.  In Figure
\ref{fig:sugra4}  we plot the same information for 
\tanb\  = 40, \azero\ = 0.  We see here about half the parameter space can be
scanned by the CDF detector  (and the whole parameter space if a similar
analysis holds for the D0 detector).  See Figure \ref{fig:bsmumu_br_limits}.  We
note, further, that if \azero\ = 0,  a simultaneous measurement  of both
$Br[\bsmumu]$ and $a_\mu$ would essentially determine the mSUGRA  parameters, as
the \mzero\ allowed region at fixed \mhalf\
 is very narrow due to  the dark matter constraint.  The effect of varying
\azero\ is shown in Figure \ref{fig:sugra5},  where the allowed region for 
$\azero\  = -2 \mhalf$, \tanb\ = 40 in the \mzero-\mhalf\  plane is plotted. The
effect is to tilt (and narrow) the allowed dark  matter band. The entire allowed
parameter space can again be probed.  

In Figures \ref{fig:sugra3}, \ref{fig:sugra4} and \ref{fig:sugra5}, we have also drawn  lines for
various light Higgs masses (vertical dotted lines). A measurement of $\bsmumu$,
$a_{\mu}$ and $m_h$ would then effectively determine the parameters of mSUGRA
for $\mu>0$ by requiring that they intersect with the dark matter allowed band 
at a point. (If no choice of parameters allowed this, mSUGRA would be ruled
out.) The Tevatron Run II should be able to either rule out a Higgs mass or give
evidence for its existence at the 3
$\sigma$ level over the entire allowed mass range of SUSY light Higgs masses.
Alternatively, the LHC's determination of $m_h$ or the gluino mass (to determine
$m_{1/2}$) would fix the parameters of mSUGRA.

The BNL E821 experiment should shortly have a more accurate value  for $a_\mu$, 
with errors reduced by a factor of two or more. We note from the above  figures
the importance this result might have.  Thus if $a_\mu$ increases, the 
$a_\mu$ bound moves downward, encroaching further on the allowed part of the 
parameter space, and a value of 
$a_\mu \,\gtsim\ 50 \times 10^{-10}$ would eliminate the  mSUGRA model \cite{Arnowitt:2001be}. 
However, if $a_\mu$ decreases significantly (but is still  positive), the mSUGRA
model would predict a heavy SUSY particle spectra  closer to the TeV region,
having significant effects on accelerator and  dark matter detection physics. An
accurate determination of $a_\mu$  corresponds to a line from upper left to
lower right (or more precisely a  band when errors are included) running
parallel to the 
$a_\mu < 11 \times 10^{-10}$  boundary, and cutting through the allowed dark
matter band which runs from  lower left to upper right. Thus these two
experiments are complementary for  determining the mSUGRA parameters.


\noindent {\bf 6. R PARITY VIOLATING MODELS}. We consider briefly here the case when R parity is
broken. The general  renormalizable  R parity violating superpotential has the
form
\begin{equation}
	W_{\rpv} = \kappa_iL_iH_2 + \lambda_{ijk} L_iL_jE^c_k
	+ \lambda'_{ijk} L_iQ_jD^c_k
	+ \lambda''_{ijk}U^c_iD^c_jD^c_k
\end{equation} The first three terms are lepton number violating, and the last
term is  baryon number violating. For the \bsmumu\ decay, we need to consider 
lepton violating terms\cite{jang}, and so we set $\lambda^{\prime\prime}$
  to zero (to prevent rapid  proton decay). Unlike the R parity conserving
models, the SUSY contribution  to \bsmumu\ can now occur at the tree level, and
so can be considerably  larger. For example, if the $\lambda$ and $\lambda'$
terms are present in the  theory, we can have
\begin{equation} 
 C_S=-{{\lambda^{\prime}_{i23}\lambda^*_{i22}}\over{2m_{\tilde\nu}^2}},\,\, 
 C_P={{\lambda^{\prime}_{i23}\lambda^*_{i22}}\over{2m_{\tilde\nu}^2}}.
\end{equation}

In Figure \ref{fig:rpv1},  we plot the $Br$ as a function of $\mhalf$ 
 in the $R$ parity violating scenarios.
 We have chosen 
$\lambda^{\prime}_{i23}=\lambda_{i22}=0.02$ and $\tan\beta=10$.  These small
couplings are not restricted by any another physical process\cite{rpv}.  We see
that in this case smaller values of $\tanb$  can be probed and the $Br$ is large
compared to the R parity conserving case.  The contributions of $C_{S,P}$ to the
newly discovered decay mode
$B_s\rightarrow K\mu\mu$ are small. For example, for 
$\mhalf = 300$ GeV and $\mzero = 300$ GeV,  $Br[B_s\rightarrow K\mu\mu]$ gets a
contribution of $3.30\times 10^{-7}$.  (This calculation, however, involves a
large QCD uncertainty.) The observed branching fraction for this mode is
$(0.99^{+0.4+0.13}_{-0.3-0.14}) \times 10^{-6}$\cite{Abe:2001dh}.

In contrast to the R conserving case, the R parity violating scenarios are
not infested with tau's and therefore has the potential to be
 observed at Run II directly.  For example, let us consider the production
 of $\chi_1^\pm\chi^0_2$ and let us assume that 
 $\lambda_{i22},\,\lambda'_{i23}$ are
 present. In this case we can have several interesting final states  such as
$6l+2$jets+missing $E_T$ (from
 neutrino) and  $4l\,+\,4$jet+missing $E_T$ from the production of
$\tilde\chi^{\pm}_1$-$\tilde\chi^0_2$. 


\noindent{\bf 7. CONCLUSIONS}. We have investigated the prospects for studying the rare  decay
mode \bsmumu\ by the CDF detector in Run II at the Tevatron. We  have analyzed
the background for this process and find that a 
$Br > 1.2 \times 10^{-8}$ for 15 fb$^{-1}$
 can be probed at Run II. (This is nearly a factor of  100 improvement over the
Run I bound.) In this background analysis, we have  not only extrapolated the
Run I limit, but also introduced new selection  cuts to reduce the background.
The improvement of the Run II detector has  also been taken into account. 
If a similar analysis can be  performed for the D0 detector, the
combined data would
increase the sensitivity for  detecting this decay mode to $Br>6.5\times
10^{-9}$. The LHC should be able to reach the sensitivity at the level of SM
branching ratio\cite{Nikitenko:1999ak}.

The \bsmumu\ is an important process for SUSY searches for new physics,  as the
Standard Model prediction of the branching ratio is quite small  ($3.5 \times
10^{-9}$),  and the SUSY contribution increases for large $\tan\beta$ as 
$\tan^{6}\beta$.  For the mSUGRA model, the above sensitivity implies that Run
II  could probe a region of parameter space for $\tanb > 30$, a region which 
could not be probed by a direct search at Run II.  We have combined the 
expectations for \bsmumu\  for mSUGRA with other experimental bounds on  the
parameter space. Thus a large branching ratio, 
\ie,  $> 7 \times 10^{-8} (14 \times 10^{-8})$ would  be sufficient to 
eliminate the mSUGRA model for $\tan\beta\leq 50(55)$. 
A measurement of \bsmumu\ , the muon $g -2$ anomaly and the light Higgs mass combined with the
astronomical bounds on cold  dark matter would essentially determine the mSUGRA model,
allowing predictions of  all other sparticle masses and cold dark matter
neutralino-proton  cross  sections. If the muon $g - 2$ anomaly remains positive
these cross sections  are $\gtsim 10^{-9}$ pb,  and should then be accessible to
future planned  detectors such as GENIUS, Cryoarray, ZEPLIN IV and CUORE.

A simple model of R parity violation was also considered, and here the 
$Br[\bsmumu]$ can be large for both small and large $\tan\beta$.


\noindent{\bf ACKNOWLEDGEMENTS} This work was supported in part by National Science Foundation grant 
PHY-0101015 and  in part by Department of Energy grant DE-FG03-95ER40917.  We
should  like to thank D.~Toback for valuable comments.

\renewcommand{\baselinestretch}{1}

\newpage
\begin{figure}
\centerline{ \DESepsf(figadkm.epsf width 8 cm) }
\caption {\label{fig1}  Example of diagram contributing to \bsmumu\ with leading 
contribution of $\tan^{3}\beta$.  The $H$ and $A$ are the heavy CP even and CP
odd  neutral Higgs bosons, $\schi^{\pm}_{i}$ ($i$ = 1, 2) are charginos and
$\tilde t_i(i=1,2)$ are the stop bosons. 
Heavy marked vertices each contain a factor of $\tan\beta$.}
\end{figure}
\vspace{-9cm}

\begin{figure}
\begin{center}
    \epsfig{file=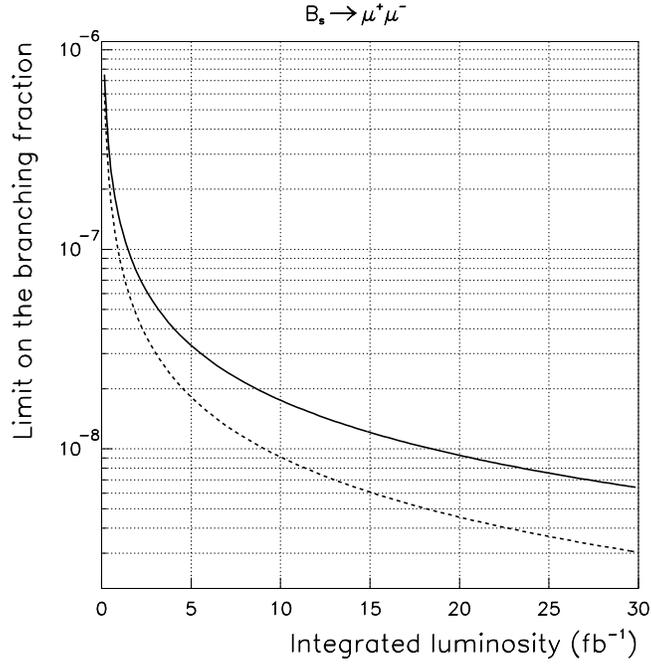,height=250pt}
    \caption{Illustrated 95\% C.L. limits on the branching ratio for
	$B_s \to \mpmm$ at CDF in Run II as a function of
	integrated luminosity.
	Solid (Case A) and dashed (Case B) 
	curves are based on different assumptions
	on the signal selection efficiency and the background
	rejection power. See the text for details.}
    \label{fig:bsmumu_br_limits}
\end{center}
\end{figure}
\begin{figure}\vspace{-2cm}
    \begin{center}
    \leavevmode
    \epsfysize=7.0cm
    \epsffile[75 160 575 630]{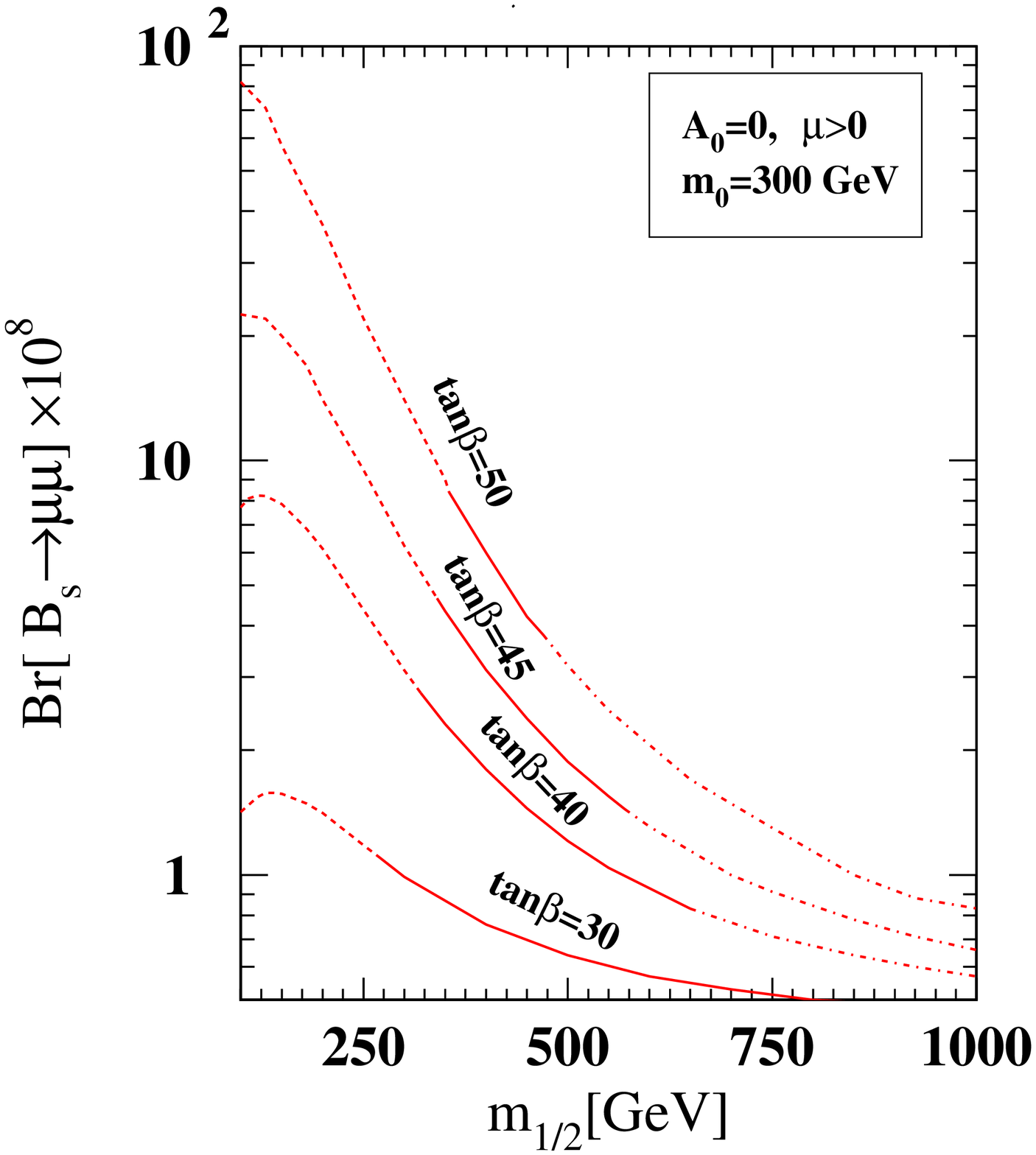} 
    \vspace{2.0cm}
    \caption{\label{fig:sugra1} Branching ratio for
$B_s \to \mpmm$  as a function of  $\mhalf$  for various $\tanb$ values in
mSUGRA models. Other mSUGRA parameters are fixed to be $\mzero$ = 300 GeV,
$\azero = 0$ and $\mu > 0$. Dashed and dash-dotted lines are to indicate the
models are excluded via constraints on $Br[b \to s \gamma]$ 
and $m_{\stau} > m_{\lsp}$, respectively.}
    \end{center}
\end{figure}

\begin{figure}\vspace{-2cm}
    \begin{center}
    \leavevmode
    \epsfysize=7.0cm
    \epsffile[75 160 575 630]{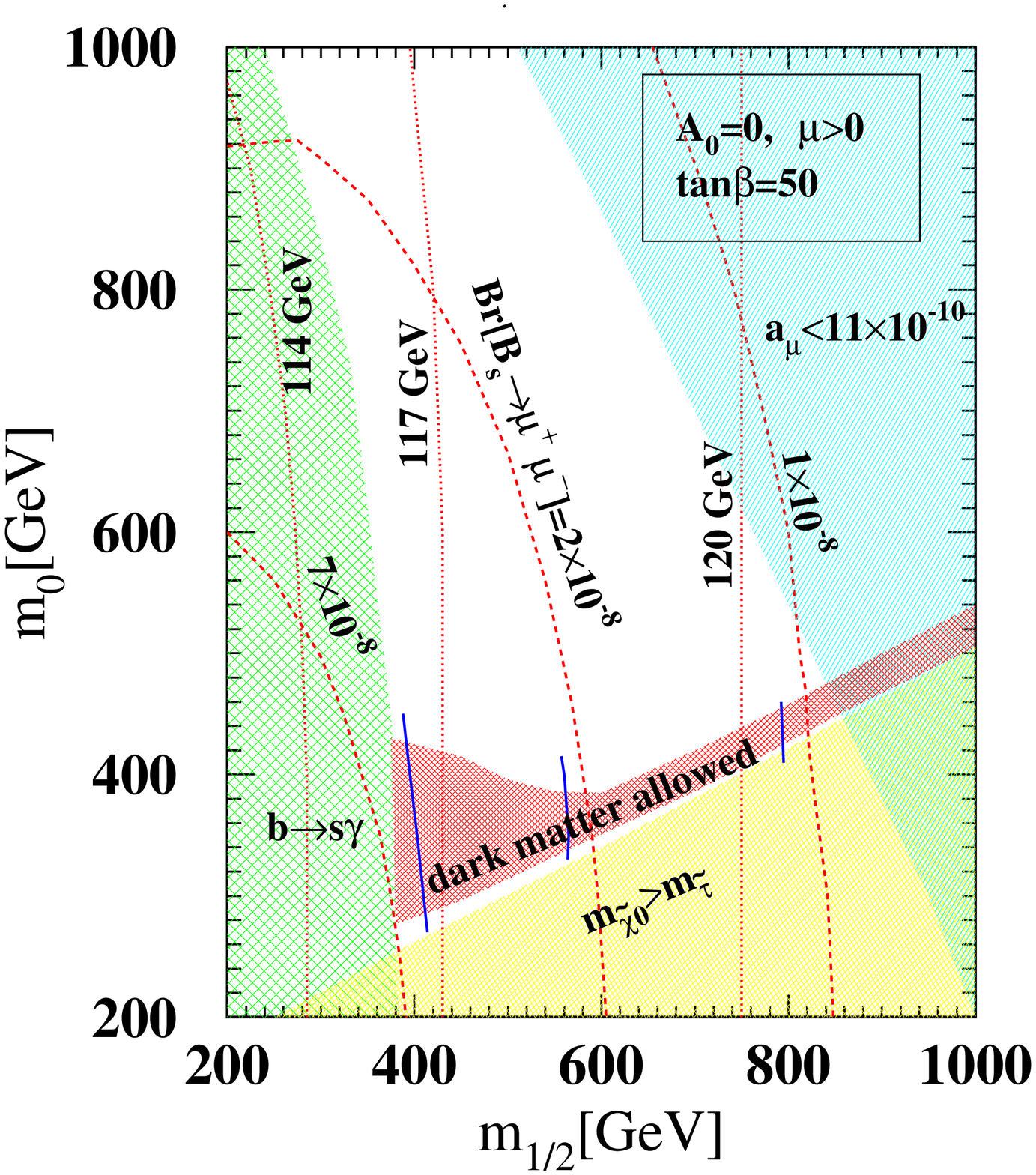} 
    \vspace{2.0cm}
    \caption{\label{fig:sugra3} Branching ratio for $B_s \to \mpmm$ (three dashed lines from left
to right: 
$7\times 10^{-8}$, $2\times 10^{-8}$, $1\times 10^{-8}$) for $\tanb$ = 50 in the
$\mzero$-$\mhalf$ plane. Other mSUGRA parameters are fixed to be 
$\azero = 0$ and $\mu > 0$.
 The three short solid lines indicate the $\sigma_{\tilde\chi^0_1-p}$ values 
(from left:  0.05 $\times 10^{-6}$ pb,  0.004 $\times 10^{-6}$ pb, 0.002 $\times
10^{-6}$ pb).  The vertical dotted lines label Higgs masses.}
\end{center}
\end{figure}

\begin{figure}\vspace{-2cm}
    \begin{center}
    \leavevmode
    \epsfysize=7.0cm
    \epsffile[75 160 575 630]{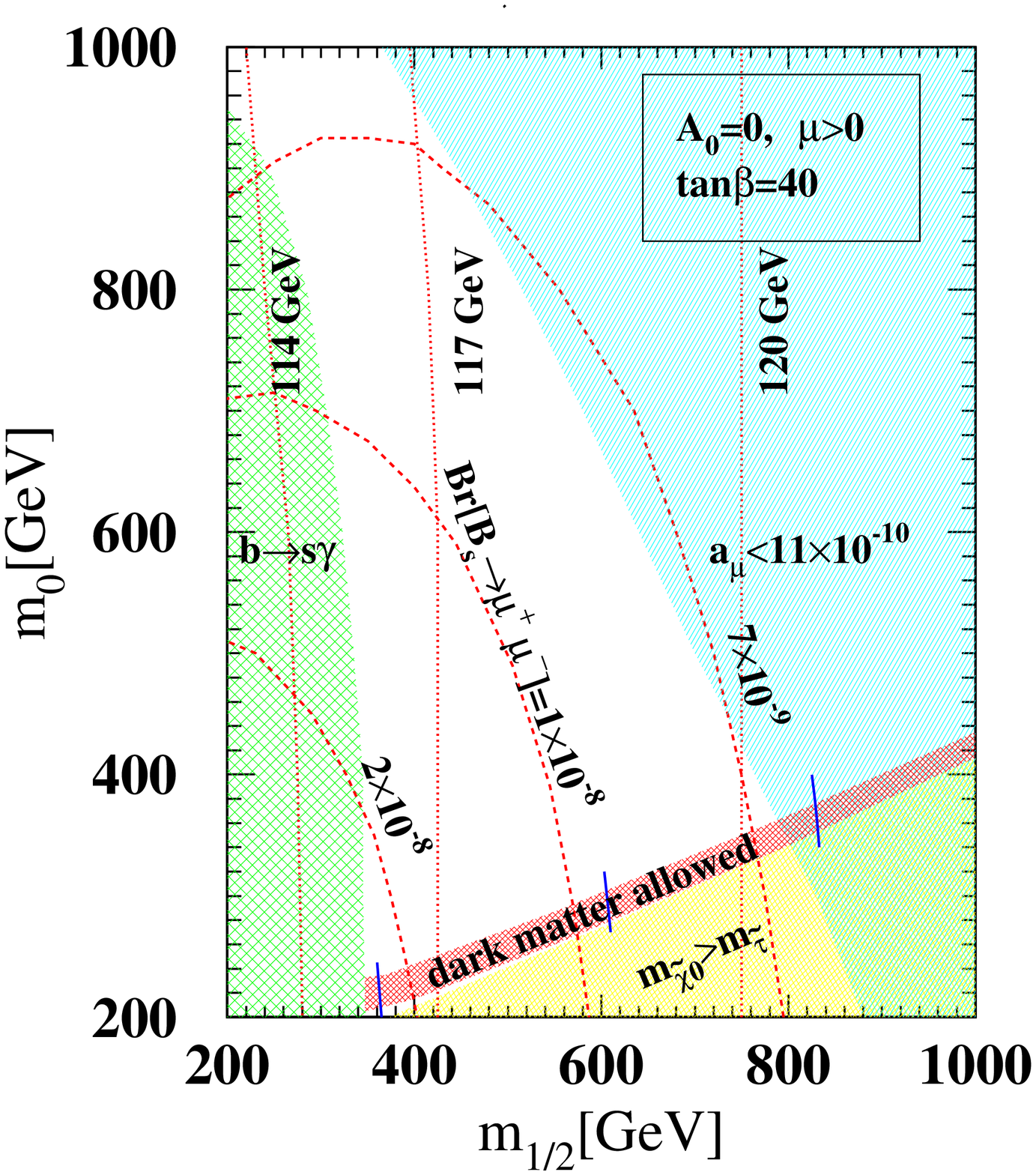} 
    \vspace{2.0cm}
    \caption{\label{fig:sugra4} Branching ratio for $B_s \to \mpmm$  (three dashed lines from left
to right: $1.9\times 10^{-8}$, $1\times 10^{-8}$, 
$0.7\times 10^{-8}$) at $\tanb$ = 40 in the $\mzero$-$\mhalf$ plane. Other
mSUGRA parameters are fixed to be
$\azero = 0$ and $\mu > 0$.
 The three short solid lines indicate the $\sigma_{\tilde\chi^0_1-p}$ values 
(from left: 0.03 $\times 10^{-6}$ pb,  0.002 $\times 10^{-6}$ pb, 0.001 $\times
10^{-6}$ pb).  The vertical dotted lines label Higgs masses.}
\end{center}
\end{figure}

\begin{figure}\vspace{-2cm}
    \begin{center}
    \leavevmode
    \epsfysize=7.0cm
    \epsffile[75 160 575 630]{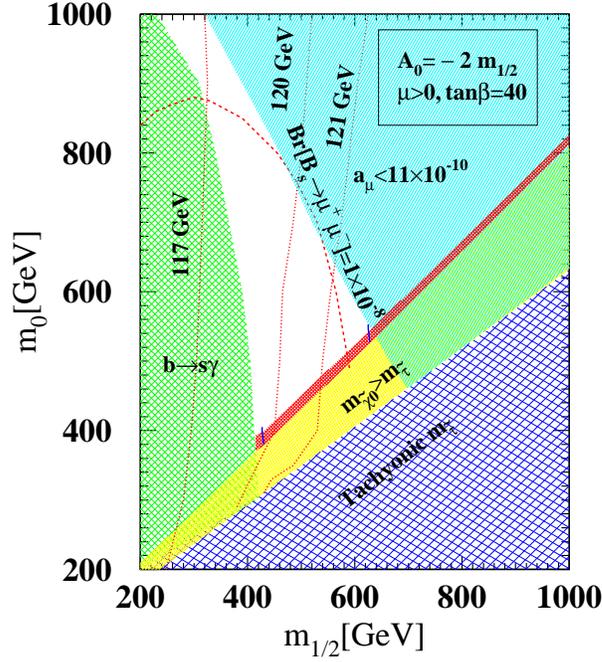} 
    \vspace{2.0cm}
    \caption{\label{fig:sugra5}Branching ratio for $B_s \to \mpmm$  at $\tanb$ = 40 in the
$\mzero$-$\mhalf$ plane for $A_0=-2m_{1/2}$
 and $\mu > 0$. The two short solid lines indicate the
$\sigma_{\tilde\chi^0_1-p}$ values (from left: 
 0.005$\times 10^{-6}$ pb, 0.001 $\times 10^{-6}$ pb). 
 The vertical dotted lines label Higgs masses.}
\end{center}
\end{figure}

\begin{figure}\vspace{-2cm}
    \begin{center}
    \leavevmode
    \epsfysize=7.0cm
    \epsffile[75 160 575 630]{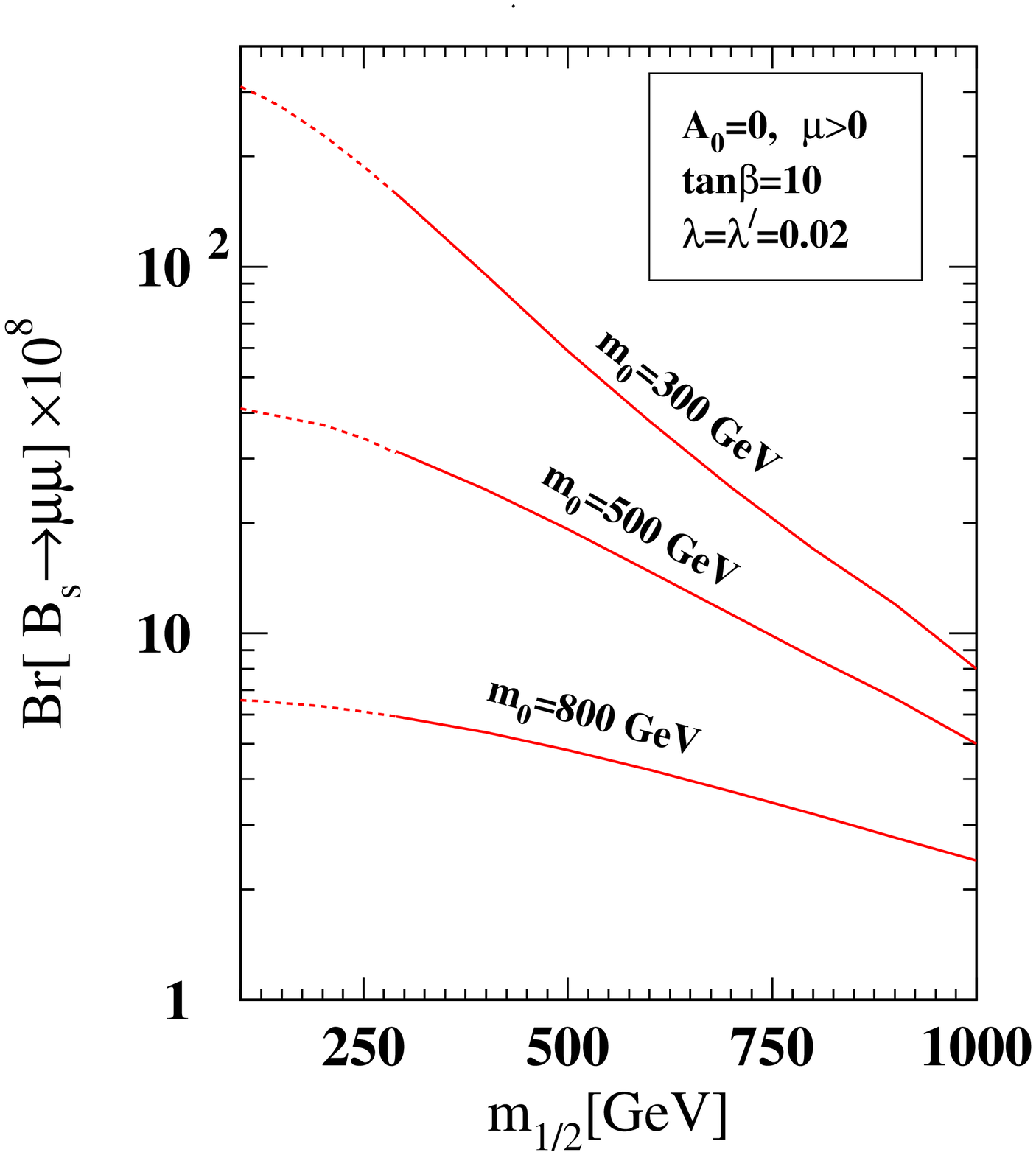} 
    \vspace{2.0cm}
    \caption{\label{fig:rpv1}Branching ratio for
$B_s \to \mpmm$  as a function of  $\mhalf$ (in GeV) for $\mzero$ = 300, 500,
and 800 \gev  in a \rpv\ SUSY scenario ($\lambda_{i22} = \lambda^{\prime}_{i32}$
= 0.02). Other mSUGRA parameters are fixed to be $\tanb$ = 10,
$\azero = 0$ and $\mu > 0$. Dashed lines are to indicate the models that 
are excluded
via the $b\rightarrow s\gamma$ constraints.}
\end{center}
\end{figure}


\begin{thebibliography}{99}
\bibitem{def} We use a coordinate system where $\theta$ and $\phi$ are the polar and azimuthal
angles, respectively, with respect to the proton beam direction ($z$ axis). The pseudorapidity
($\eta$) is defined as $-ln[\tan(\theta/2)]$. The transverse momentum of a particle is denoted as
$p_T=p\,sin\theta$.
\end{thebibliography}
\end{document}